\documentclass[aps,prb,twocolumn,superscriptaddress]{revtex4-2}
\usepackage{bm}
\usepackage{graphicx}
\usepackage{graphics}
\usepackage{amsmath}
\usepackage{amsfonts}
\usepackage{amssymb}
\usepackage{makeidx}
\usepackage[colorlinks=true,citecolor=blue,linkcolor=blue,linktocpage=true,pagebackref=false]{hyperref}
\hypersetup{colorlinks=true,citecolor=blue,linkcolor=blue,filecolor=blue,urlcolor=blue}
\usepackage{color,soul} 
\usepackage{float}
\usepackage{subfigure}
\usepackage{xcolor}

\usepackage[normalem]{ulem} % permite usar fonte tachada

\usepackage{IEEEtrantools}

\newcommand{\red}[1]{\textcolor{red}{ {#1}}}

\graphicspath{{figures/}} 
\begin{document}
\newcommand{\be}   {\begin{equation}}
\newcommand{\ee}   {\end{equation}}
\newcommand{\ba}   {\begin{eqnarray}}
\newcommand{\ea}   {\end{eqnarray}}
\newcommand{\ve}  {\varepsilon}
\newcommand{\Dis} {\mbox{\scriptsize dis}}
\newcommand{\state} {\mbox{\scriptsize state}}
\newcommand{\band} {\mbox{\scriptsize band}}
%%%%%%%%%%%%%%%%%%%%%%%%%%%%%%%%%%%%%%%%%%%%%%%%%%%%%%
\title{Driving electronic features of twisted bilayer zigzag-graphene nanoribbons\\
} 

\author{Kevin J. U. Vidarte}
\affiliation{Instituto de F\'{\i}sica, Universidade Federal
Fluminense, 24210-346 Niter\'oi, RJ, Brazil}
\author{A. B. Felix}
\affiliation{Instituto de F\'{\i}sica, Universidade Federal
Fluminense, 24210-346 Niter\'oi, RJ, Brazil}
\author{A. Latg\'e}
\affiliation{Instituto de F\'{\i}sica, Universidade Federal
Fluminense, 24210-346 Niter\'oi, RJ, Brazil}

\date{\today}
%----------------------------------------------------------------
\begin{abstract}
Novel physical properties have been reported recently by stacking graphene-like systems in different configurations. Here, we explore the nature of emergent localized states at the edges of twisted bilayer graphene nanoribbons.  Based on an extended tight-binding Hamiltonian, which includes hopping energy within a wide atomic neighborhood, we investigate the nature of the electronic states responsible for the transport along the four graphene nanoribbon terminals. The emphasis is on discussing the role of the stacking region symmetries, the twisted angle between the crossed zigzag nanoribbons, and also the width of the ribbons in the electronic and transport responses of the four terminals. Our findings show a direct connection between the number of non-equivalent sites on the edge of the stacking region and the localized states, in accordance with reported scanning tunneling spectroscopy measurements. Within the parameters explored, the twist angle was revealed to be the most powerful tool to control transport responses in the investigated 4-terminal devices, including special electronic beam splitter phenomena.

\end{abstract}
%------------------------------------------------------------------------
\maketitle

%%%%%%%%%%%%%%%%%%%%%%%%%%%%%%%%%%%%%%
\section{Introduction}
\label{Sec:introduction}
%%%%%%%%%%%%%%%%%%%%%%%%%%%%%%%%%%%%%%
Graphene nanoribbons emerged providing a large number of promising physical features, such as half-metallicity under applied electric fields \cite{Cohen2006,Dutta2009}, spin-polarized currents when corrugated \cite{Leonor2020} or doped on edges \cite{Nature2021}, and good thermoelectric properties when in semiconducting configurations \cite{Ouyang2009, Lage2022,Kim2025}.
On-surface synthesis has been successfully applied to fabricate many different armchair graphene nanoribbons (AGNR) \cite{Cai2010,armchairNR}.
Bottom-up synthesis of zigzag graphene-nanoribbons (ZGNRs) through surface-assisted polymerization and cyclodehydrogenation succeeded in yielding atomically precise zigzag edges \cite{ZIGZAGribbons}.
The favorable experimental advances have allowed the design of different scenarios for composing graphene structures to be used as electronic devices with high capacity to handle the electronic states.

%--------------------------- F I G U R E ----------------------------------------
\begin{figure}[h!]
\centering
\includegraphics[width=0.85\linewidth]{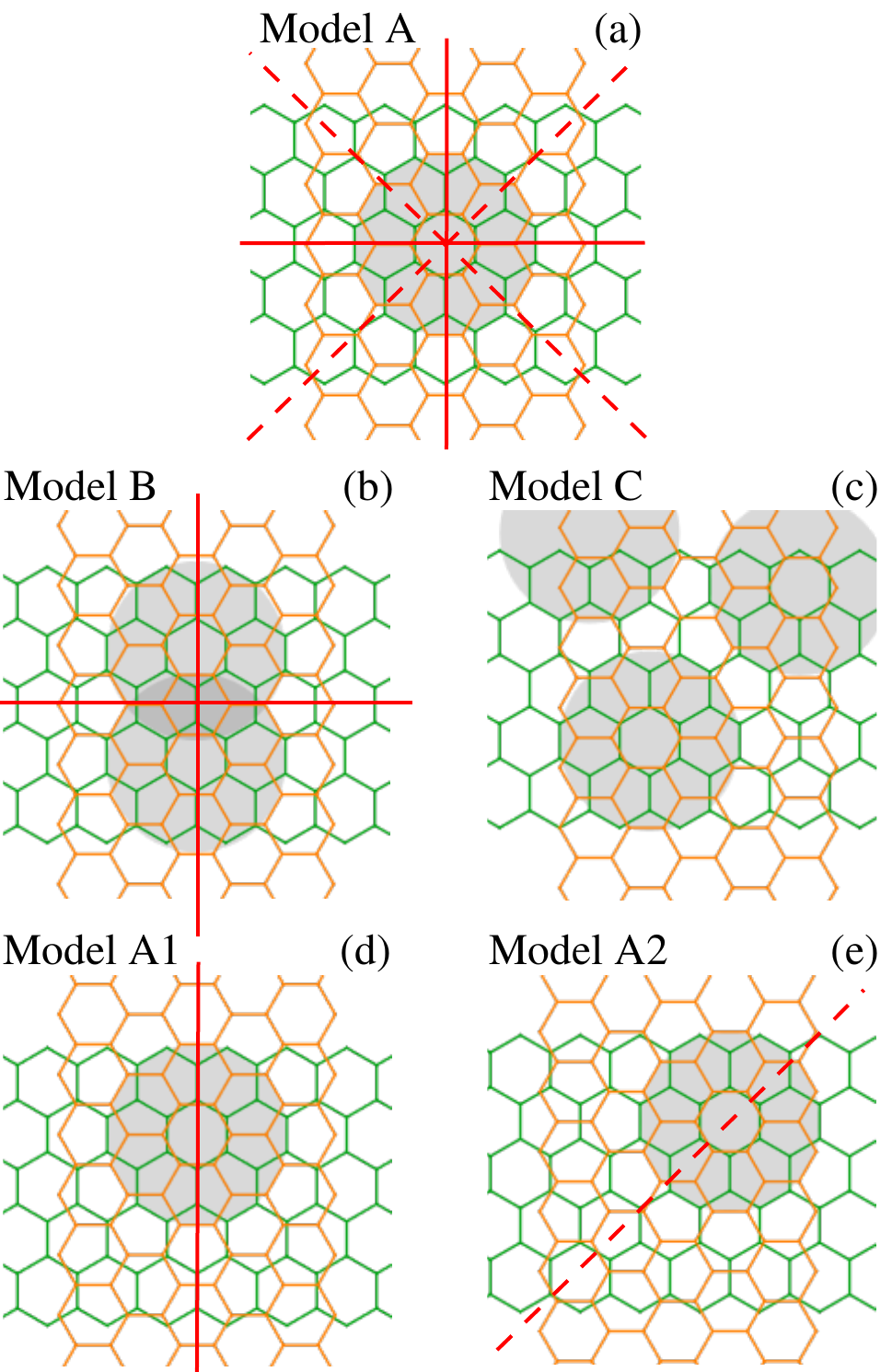}
\caption{Atomic structures of two crossed 6-ZGNRs with a relative angle $\theta = 90^{\circ}$, corresponding to five stacking configurations (a-e), named as A, B, C, A1, and A2 models, respectively.
Green and orange colors define the bottom and top nanoribbons.
Red dashed and solid lines mark the $C_{2}$ symmetry and vertical mirror, respectively.
Gray circular areas highlight 12-fold local-symmetry regions.}
\label{fig:Fig_01}
\end{figure}
%-------------------------------------------------------------------------------

Twisted bilayer zigzag-graphene nanoribbon junctions (TBZGNRs) have recently been constructed and characterized on a surface of Au (111) using scanning tunneling microscopy \cite{Wang2023}.
The TBZGNR devices were designed with a twisted angle $\theta$, well controlled by lateral manipulation of the STM tip, and $\theta$ ranging from 30$^{\circ}$ to 90$^\circ$.
Based on first-principles calculations and experimental observations, the emergent edge states in the TBZGNR stacking region are found to be highly tunable, revealing that, in addition to the twist angle, which is the main factor in 2D cases, the dominant role is the in-plane stacking offset in twisted bilayer 1D systems.
Highly tunable edge states have been proven to emerge in particular configurations of the stacked nanoribbons. The transport properties of a bilayer twisted graphene flake contacted by two monolayer nanoribbons were studied, revealing that the low energy transport properties are governed by the edge states with AB stacking\cite{Leo2015}. 
Other studies on crossed graphene nanoribbons have been reported \cite{PhysRevB.102.035436,Spin2022,Brandimarte2017,Caio2016}, which predict beam splitters and electron mirrors when coupled to nanodevices. Most of them restricted the analysis within the particular AA and AB stacking configuration of crossed graphene nanoribbons with all edges in the zigzag or armchair configurations. 
Small deviations from the commensurate stacking considering a twisted angle equal to $60^{\circ}$ are also reported in the four-terminal device composed of the two crossed ZGNRs \cite{Sanz2024}. 

In this paper, we discuss distinct stacking possibilities for TBZGNR systems and analyze the highly tunable edge states more generally for sliding displacements in the plane, varying the twist angle and the ribbon widths in crossed ZGRNs. Our analysis is based on the non-equivalent sites of the edges of the stacking region defined by the system symmetry and size of the stacking region. 
A Slater Koster tight-binding approximation is adopted and we use the HHK (Haydock, Heine, and Kelly) model \cite{Haydock1972,Haydock1980,Vidarte2022} to study the electronic structure of the systems based on the local atomic environment.
Transport calculations are addressed via Landauer-B\"uttiker formula following the Green function formalism \cite{Data}.

%%%%%%%%%%%%%%%%%%%%%%%%%%%%%%%%%%%%%%
\section{Theory}
\label{Sec:Theory&method}
%%%%%%%%%%%%%%%%%%%%%%%%%%%%%%%%%%%%%%

\subsection{TBZGNRs: geometric and symmetry properties}
\label{Subsec:90-TBZGN&symmetry}
The stacking geometry of the two crossed ZGNRs is defined by starting from two AA-stacked ZGNRs and subsequently rotating the top ZGNR by an angle $\theta$ around the hexagonal center shared of both nanoribbons, as shown in Fig.~\ref{fig:Fig_01}. 
Different configurations may be achieved by sliding the top ribbon with respect to the bottom nanoribbon. 

For $\theta = 90^{\circ}$, different 6-TBZGNR models of the in-plane stacking offset are considered; 6 being the number of carbon atom chains composing the ZGNR.
Model A presents a shared hexagonal center of both nanoribbons, located between the third and fourth zigzag chains of carbon atoms, as shown in Fig.~\ref{fig:Fig_01}(a).
This model has two mirrors (red lines) and two $C_{2}$ symmetries (red dashed lines), adopting the TBZGNR system with high symmetry.
Model B, presented in Fig.~\ref{fig:Fig_01}(b), is obtained by a lateral translation of the bottom nanoribbon by $\pm \sqrt{3}a_0/2 \hat{\bf e}_{x}$ starting from Model A. Note that the symmetry $C_{2}$ is absent in this new configuration.
Fig.~\ref{fig:Fig_01}(c) displays Model C, obtained by the lateral translations of $4a_0 \hat{\bf e}_{x}$ and $-4a_0 \hat{\bf e}_{y}$ in the bottom and top nanoribbons, respectively, from Model A. No lattice symmetry is predicted in Model C. These three models were proven and analyzed in detail by scanning tunneling spectroscopy, as reported in Ref.~\cite{Wang2023}.

The stacking area in different 6-TBZGNR models corresponds to the portion of quasicrystalline $30^{\circ}$ twisted bilayer graphene \cite{Yao2018,Ahn2018,Yan2019,Vidarte2024Q,Moon2019,Yu2019,Stampfli1986}.
Fig.~\ref{fig:Fig_01} shows circular gray areas that indicate the local symmetry regions associated with 12-atom rings that, in two dimensions, exhibit a characteristic dodecagonal quasi-crystal pattern \cite{Vidarte2024Q,Yan2019}.
Contrary to the definition ``moir\'e sites with AA-stacking'', used in Ref.~\cite{Wang2023},  here we named it as ``12-fold local-symmetry regions''.
For example, Model A exhibits a perfect 12-fold symmetry, whereas Models B and C exhibit other local symmetry regions related to Stampli tiles \cite{Vidarte2024Q,Stampfli1986}. The perfect 12-fold symmetry always occurs on the axis of rotation. We then propose two additional models with perfect 12-fold symmetry in particular local-symmetry regions.
In Model A1, as shown in Fig.~\ref{fig:Fig_01}(d), the local symmetry region is situated on the top of the bottom nanoribbon and has only one mirror axis, while in Model A2 it is situated at one corner of the stacking area and presents only one $C_{2}$ symmetry, as shown in Fig.~\ref{fig:Fig_01}(e).
The symmetries of each model have a direct influence on the number of non-equivalent sites at the edges of the stacking region.

\subsection{Model Hamiltonian}
\label{Subsec:ModelHamiltonian}

\subsubsection{Tight-binding model}

The low-energy electronic properties of TBZGNRs are obtained using a tight-binding model defined on an orthogonal basis as \cite{Castro2009,Reich2002}
\be
\label{Eq:Tight-binding}
H=\sum_{i,j}t({\rm \textbf{R}}_{i} - {\rm \textbf{R}}_{j}) \vert {\rm \textbf{R}}_{i}\rangle \langle {\rm \textbf{R}}_{j} \vert + {\rm h.c.},
\ee
where $\langle {\rm \textbf{r}}\vert {\rm \textbf{R}}_{i}\rangle$ is the corresponding Wannier $p_{z}$ electronic orbital centered at the $i$th atom at the position ${\rm \textbf{R}}_{i}$. The sum in the Hamiltonian involves all the i and j sites of the TBZGNR.
The transfer integral $t({\rm \textbf{R}})$ depends on the interatomic distance and the relative orientation between $p_{z}$ orbitals connected by ${\rm \textbf{R}}$.
The transfer integral is parameterized following the Slater-Koster model as \cite{Slater1954}
\begin{equation}
-t({\mathbf R}) = V_{pp\pi} (R)  \left[ 1-\left( \frac{ {\mathbf R} \cdot  \mathbf{e}_z } {R} \right)^2 \right] +  V_{pp\sigma} (R) \left( \frac{ {\mathbf R} \cdot  \mathbf{e}_z } {R} \right)^2,
\end{equation}
where the hopping parameters $V_{pp\pi}(R)$ and $V_{pp\sigma} (R)$ are given by the exponential decay functions,
\begin{equation}
 V_{pp\pi(\sigma)} (R) =  V^0_{pp\pi(\sigma)} \exp \left( -\frac{R - a_0(d_0)}{ r_0} \right) 
\end{equation}
with  $V^0_{pp\pi} = -2.65$ eV and $V^0_{pp\sigma} = 0.48$ eV, $r_0= 0.184 \, a$ is the decay length of the hopping matrix element \cite{Trambly2010,Moon2013,Uryu2004,Vidarte2024Q,vidarte2024}, $a_0 \simeq 1.42$ \AA\
is the nearest-neighbor distance in graphene, and $d_0= 3.35$ \AA\, is the inter-layer distance. Our model does not include relaxation effects. As previously reported \cite{Wang2023} out-of-plane bending and lattice distortions are considered negligible effects.

\subsubsection{Transport calculations}

Transport properties are investigated using the Landauer-B\"uttiker approach and following the Green function formalism \cite{Data}.
Each ribbon is decoupled into three parts: a central conductor and right and left leads for $\theta = 0^{\circ}$ (or top and bottom leads for $\theta \neq 0^{\circ}$) \cite{Carbon2020,Leonor2015}, defined as two semi-infinite systems, perfectly fitting to the central conductor region. The central advanced ($a$) and retarded ($r$) Green functions are given as 
\be
G^{a,r}_{c}(E)=\Big[\omega - {H_c} - \Sigma^{a,r}(E)\Big]^{-1} ,
\ee
with
\be
\Sigma^{a,r}(E)=\Sigma^{a,r}_L(E) + \Sigma^{a,r}_R(E) + \Sigma^{a,r}_T(E) + \Sigma^{a,r}_B(E) ,
\ee
where $\omega = E \pm i\eta $, $\eta$ being an infinitesimal number.
${H}_{c} $ is the tight-binding Hamiltonian of the central part, and $\Sigma^{a,r}_{\alpha} (E)$ with $\alpha = L$, $R$, $T$ and $B$ correspond to left, right, top and bottom self-energies, respectively, given by the related surface Green functions, from which the coupling matrices $\Gamma_{\alpha} (E)= i (\Sigma_{\alpha}^{r} (E) - \Sigma_{\alpha}^{a}(E))$ are obtained.
To derive the electronic conductance from terminal $\alpha$ to terminal $\beta$, $G_{\alpha\beta}(E)=2e^2/h\mathcal{T_{\alpha\beta}}(E)$, we calculate the energy-dependent transmission given by,
\begin{equation}
\label{eq:Transmission}
{\mathcal{T}_{\alpha \beta}(E)}=\operatorname{Tr}\left[\boldsymbol\Gamma_{\alpha}\boldsymbol{G}^{r}_{c}\boldsymbol\Gamma_{\beta}\boldsymbol{G}^{a}_{c} \right]\,\,.
\end{equation}

\subsection{Numerical methodology}
\label{Subsec:NumericalMethod}

We perform a full real-space calculation of the local density of states (LDOS) for TBZGNRs using the HHK recursive method \cite{Vidarte2022,Haydock1972,Haydock1975,Haydock1980}, which is very efficient in computing large systems without periodic boundary conditions.
The HHK method proposes a very efficient Lanczos-like $O(\mathcal{N})$ recursive procedure that transforms an arbitrary sparse Hamiltonian matrix into a tri-diagonal one.
Diagonal real-space Green functions, $G^{r}_{ii}(E)\equiv G^{r}({\rm \textbf{R}}_{i},{\rm \textbf{R}}_{i},E)$, are evaluated using a continued fraction expansion, which is more amenable to numerical calculations than a complete diagonalization process.
The LDOS at any site $j$ can be written as $\rho_{j}(E)= -\frac{1}{\pi} \lim_{\eta \rightarrow 0^{+}} \left[ {\rm Im}\;G_{jj}(E+i\eta) \right]$, with a finite $\eta$ value, used as a convenient regularization parameter \citep{Vidarte2022,Haydock1975}.

For transmission coefficient calculation, we use standard recursive methods, based on decimation techniques \cite{Latge2002,Vanessa2018} to derive the surface and full Green's functions as required in Eq.~\eqref{eq:Transmission}.
This method can be used to numerically evaluate the Green functions of the system with arbitrary lattice structures and potential profiles.

%%%%%%%%%%%%%%%%%%%%%%%%%%%%%%%%%%%%%%
\section{Results and Discussions}
\label{sec:results}
%%%%%%%%%%%%%%%%%%%%%%%%%%%%%%%%%%%%%%

In this section, we present a detailed study of the single-particle density of states (DOS) focusing on the formation of low-energy bands.
The transmission probabilities between the different electrode pairs are explored and discussed. 
We divide the section into three parts.
In the first, we focus on 6-TBZGNRs with $\theta =90^{\circ}$ and analyze the change in the electronic properties of different sliding displacements in the plane.
The following subsection addresses a discussion on the role that the twisted bilayer stacking angle plays on the electronic responses of these rotated nanoribbons.
The last part explores the dependence of the electronic properties of n-TBZGNRs, described by Model A, on the width n of the crossed ZGNRs, fixing $\theta =90^{\circ}$.

\subsection{In-plane stacking offsets in crossed ZGNRs}

\begin{figure}[h!]
\centering
\includegraphics[width=0.95\linewidth]{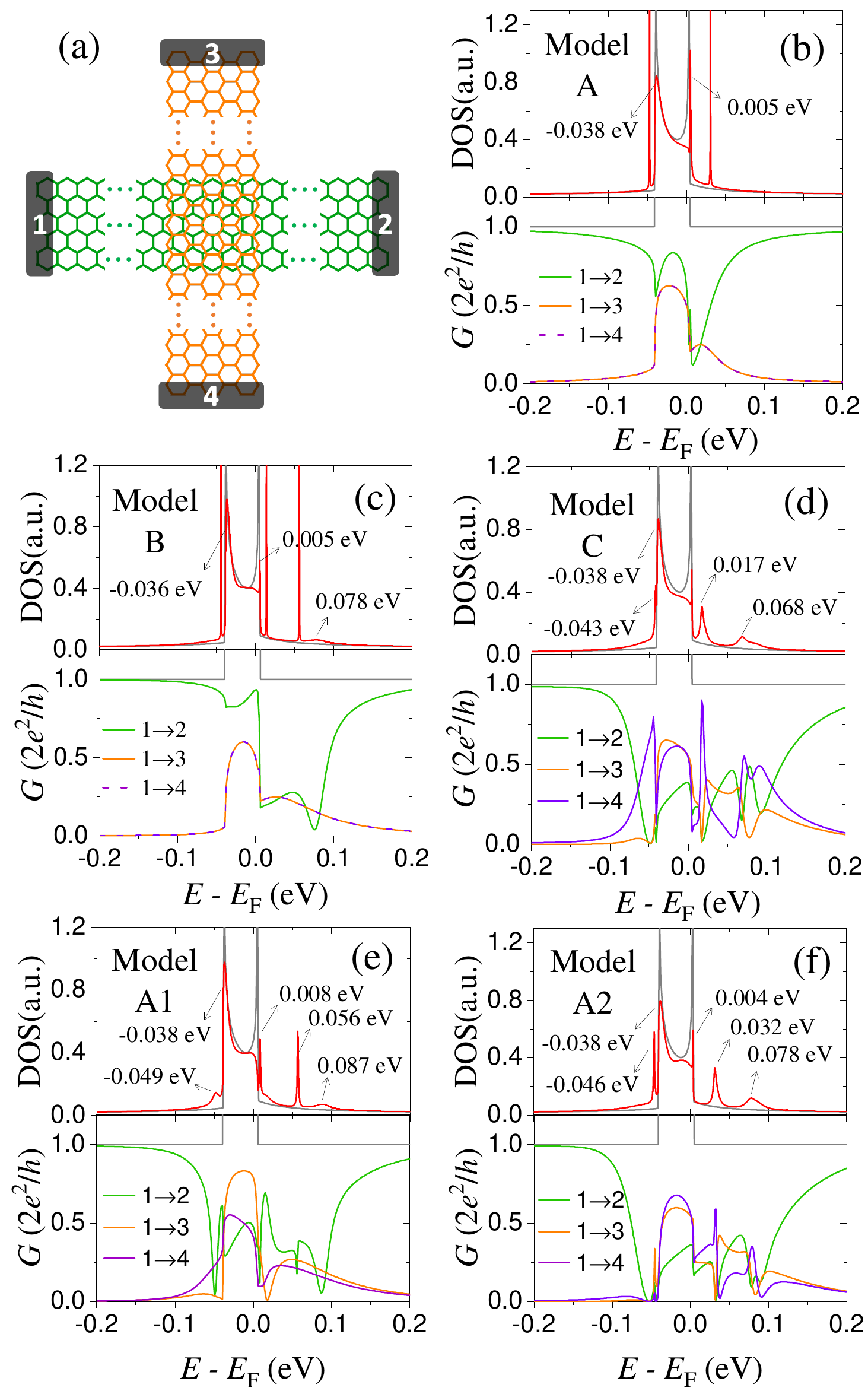}
\caption{(a) Schematic view of the proposed transport device: $90^{\circ}$ crossed zigzag-graphene nanoribbons, generating a 4-terminal device. (b-f) Density of states and conductance for the different stacking offsets of the two crossed nanoribbons. Green curves correspond to transport from lead 1 to lead 2; orange and violet denote the responses from lead 1 to lead 3 and 4, respectively. Gray curves in the panels correspond to the results of a single zigzag nanoribbon.}
\label{fig:Fig_02}
\end{figure}

Our device, shown in Fig.~\ref{fig:Fig_02}(a), contains the intersections between the two zigzag graphene nanoribbons (stacking region) and four semi-infinite electrodes described by semi-infinite zigzag graphene nanoribbons (represented as black rectangles). The left and right leads are named leads 1 and 2, while the top and bottom leads are named 3 and 4, respectively.
%--------------------------- F I G U R E ----------------------------------------

%-------------------------------------------------------------------------------

Figs~\ref{fig:Fig_02}(b)-\ref{fig:Fig_02}(f) show  DOS and conductance results around the charge neutrality point for the different 6-TBZGNR models exhibited in Fig.~\ref{fig:Fig_01}. The gray curves correspond to the results of a single 6-ZGNR, for comparison. 
By adopting the long-range electronic hopping parameterization, the flat band-like of a pristine single ZGNR at the Fermi level acquires a dispersion, splitting the characteristic high-DOS peak into two, the Fermi level being positioned between the DOS peaks.
Consequently, single 6-ZGNRs have a rectangular transmission peak around the Fermi level with a conductance value corresponding to three electron transport channels, differently from the single-channel characteristic of the first neighbor models.

%--------------------------- F I G U R E ----------------------------------------
\begin{figure}[h!]
\centering
\includegraphics[width=0.95\linewidth]{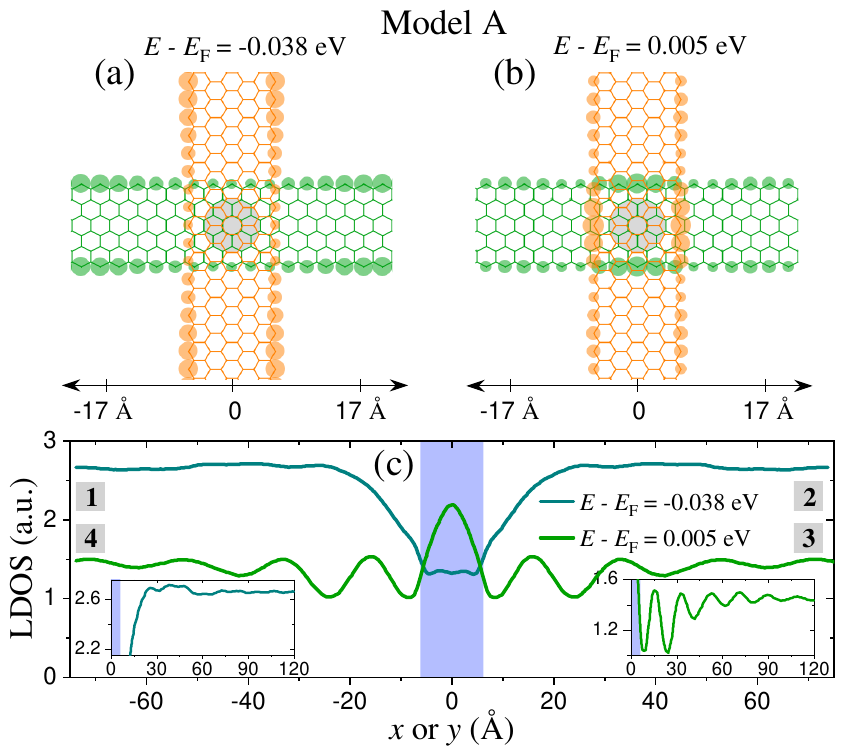}
\caption{Model A. Local density of states for $E-E_{\rm F}= -0.038$~eV (a) and $0.005$~eV (b) through the stacking region  and its next surroundings. 
(c) LDOS at the same energies at the bottom ribbon edge for both energies.}
\label{fig:Fig_03}
\end{figure}
%-------------------------------------------------------------------------------

Three emerging DOS peaks surrounding the central miniband are observed in Fig.~\ref{fig:Fig_02}(b), for the case of Model A. In contrast to the ballistic conductance of a single 6-ZGNR, the transport from lead 1 to lead 2 ($\mathcal{T}_{12}$), is greatly modified for the TBZGNR system, in the energy range of the central miniband.
Otherwise, new 1 to 3 and 4 transmission channels emerge (orange and violet curves), as illustrated in the results for $\mathcal{T}_{13}$ and $\mathcal{T}_{14}$ in Fig.~\ref{fig:Fig_02}(b).
For Model B [see Fig.~\ref{fig:Fig_02}(c)], new emerging DOS features are found in the central miniband, leading to changes in transport along the bottom ribbon ($\mathcal{T}_{12}$), when compared to the results reported in Model A.
% around $E-E_{\rm F}=0.076$~eV.}
The transmissions $\mathcal{T}_{13}$ and $\mathcal{T}_{14}$ are identical in both Models A and B, due to mirror symmetry [horizontal red solid line, in Figs.~\ref{fig:Fig_01}(a) and (b)].

%--------------------------- F I G U R E ----------------------------------------
\begin{figure}[h!]
\centering
\includegraphics[width=0.95\linewidth]{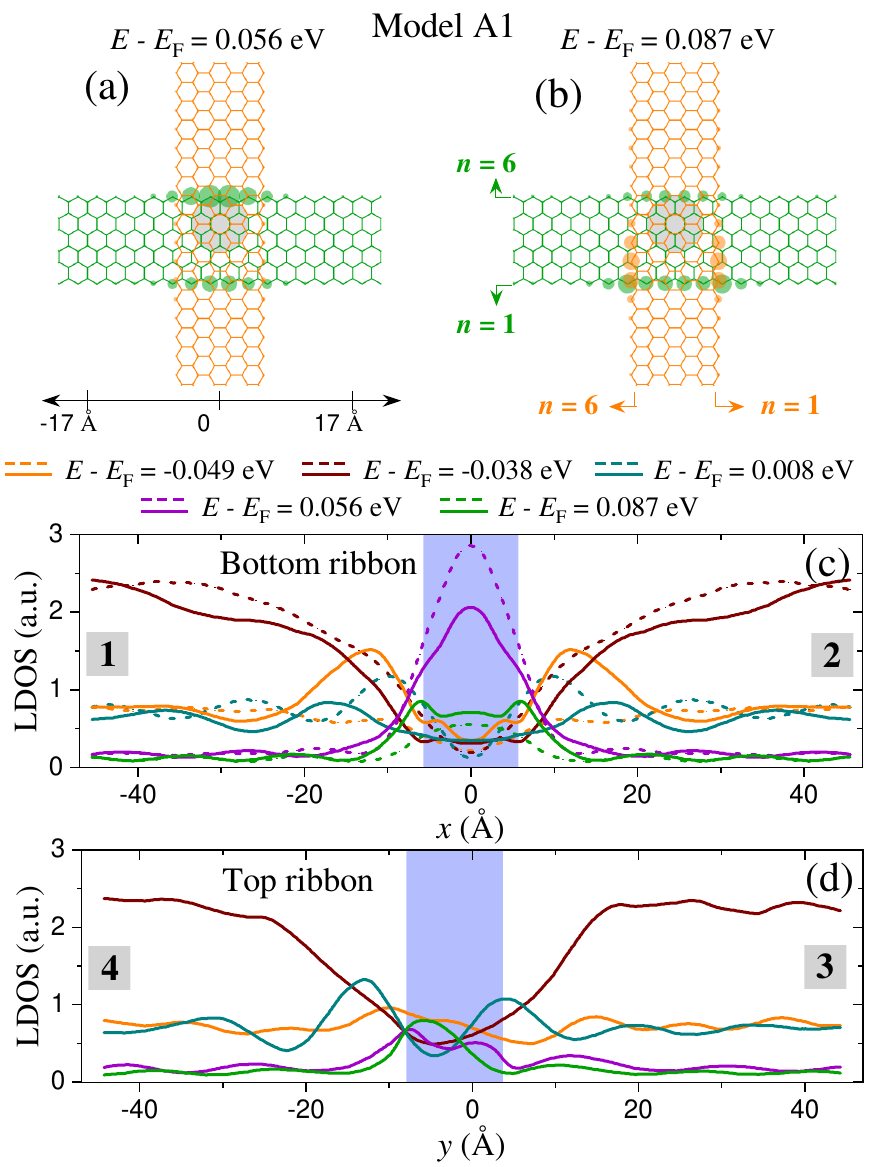}
\caption{Model A1. Local density of states at $E-E_{\rm F}= 0.056$~eV (a) and $0.087$~eV (b) through the stacking region  and its next surroundings.
LDOS at some energies at the (c) bottom and (d) top ribbon edge.
Dashed and continuous lines denotes $n=1$ and $n=6$.}
\label{fig:Fig_04}
\end{figure}
%-------------------------------------------------------------------------------

The loss of symmetry of Model C [Fig.~\ref{fig:Fig_02}(d)] leads to an anisotropic transmission from lead 1 to the upper ribbon ($\mathcal{T}_{13}$ $\ne$ $\mathcal{T}_{14}$).
The $\mathcal{T}_{12}$ also suffers a noticeable reduction in an extended energy range in the vicinity of the Fermi level.
Additional features to be mentioned are the complete transmission suppression for $\mathcal{T}_{12}$ and $\mathcal{T}_{13}$, at particular energy values, producing a filter-like effect. The less symmetric stacking configurations A1 and A2 present similar general features in the DOS and conductance compared to Model C. These stackings have been chosen to illustrate the effects that specific broken symmetries have on the transport properties responses of such carbon-basic materials. An important example is the full transport suppression, $\mathcal{T}_{12}$=$\mathcal{T}_{13}$=$\mathcal{T}_{14}=0$, involving the three leads, which is observed only for Model A2 at the threshold of the central band ($E-E_{\rm F} = -0.046 eV$).
\red{}

%--------------------------- F I G U R E ----------------------------------------
\begin{figure*} 
\centering
\includegraphics[width=0.98\linewidth]{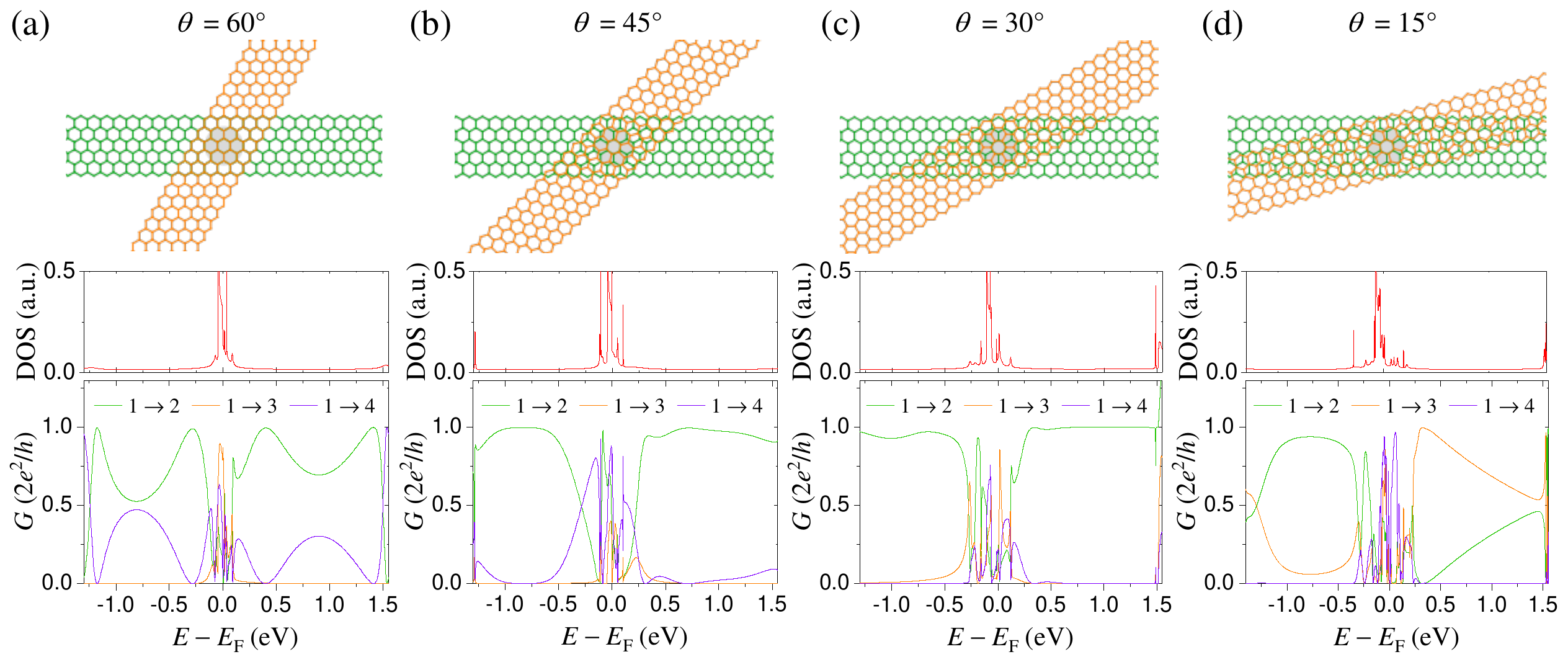}
\caption{
DOS and conductance results for angle-varying crossed zigzag graphene nanoribbons: (a) $60^{\circ}$, (b) $45^{\circ}$, (c) $30^{\circ}$, and (d) $15^{\circ}$. 
Green, orange and purples curves denote $\mathcal{T}_{12}$, $\mathcal{T}_{13}$ and $\mathcal{T}_{14}$, respectively.
Atomic structures of the 6-TBZGNRs with different angles are displayed at the top.
}
\label{fig:teste_01}
\end{figure*}
%-------------------------------------------------------------------------------
The emergence of new DOS peaks associated with localized states at the edges of the crossed nanoribbons is, in fact, a consequence of symmetry losses of the system. For the highest symmetric configuration (Model A), the main changes in transport properties occur within the flat miniband region. In the absence of $C_{2}$ symmetry (Model B), only the transmission $\mathcal{T}_{12}$ is greatly altered.
But with only one symmetry operator, for example, a horizontal mirror as in the case of Model A1, other localized states emerge (DOS peaks) affecting the transport $\mathcal{T}_{12}$ and a discrepancy between the transports $\mathcal{T}_{13}$ and $\mathcal{T}_{14}$ is raised. In the other system with only one symmetry operator presented here, Model A2, $\mathcal{T}_{13}$ and $\mathcal{T}_{14}$ exhibit also different behavior.
Finally, in the system without symmetry (Model C), the interlayer transport is clearly enhanced compared to the other proposed stacking, in the energy range analyzed.

In the following, we focus on the spatial distribution of the localized states to discuss in detail the electronic properties of the 6-TBZGNRs.
In Figs.~\ref{fig:Fig_03}(a) and (b), we show the LDOS at two energy values corresponding to the two peaks shown in the DOS of Model A [Fig.~\ref{fig:Fig_02}(a)]. The size of the green and orange symbols is proportional to the LDOS value.
The charge concentration around the edges of the stacking region is very pronounced in the case of $E-E_{\rm F} = 0.005$~eV (third peak of the central miniband). 
The details of the LDOS along the $x$- and $y$-edges are presented in Fig.~\ref{fig:Fig_03}(c) for both energies. 
The results highlight the homogeneous distribution in the case of the second peak energy ($E-E_{\rm F} = -0.038$~eV) and the preferential electronic position at the middle edges of the ZGNR system for the third peak energy.

%Similar results were found for stacking B. 
In Figs.~\ref{fig:Fig_04}(a) and (b) we present the LDOS for stacking A1 at energies outside the central miniband, corresponding to peaks in the DOS: $E-E_{\rm F}=0.056$~eV and $0.087$ eV. 
The electronic localization at the boundaries of the stacking region is evident and highlights the symmetry lost of Model A1. 
Details of the electronic distribution along the $x$- and $y$-direction of the TBZGNRs are presented in Figs.~\ref{fig:Fig_04}(c) and (d).  
While for the bottom ribbon (green ZGNR) the mirror symmetry is preserved along the $x$-axis, for the top ribbon (orange ZGNR) the $y$-axis mirror symmetry is lost. 
The LDOS for other energies corresponding to further DOS peaks are also shown. Dashed and continuous curves denote the $n=1$ and $6$ lines, as depicted in Fig.~\ref{fig:Fig_04}(b). We have also explored the electronic distribution throughout the stacking region for Model C, as discussed in Appendix~\ref{Model C}.

\subsection{TBZGNR: Varying the twist angle}

The effect of the twist angle on the electronic properties of TBZGNR is discussed here by changing the $\theta$-angle, as shown in Figs. \ref{fig:teste_01} (a)-(d).
In addition to the top views of the geometric configurations of the bilayer systems, the total density of states and corresponding conductance are also presented in Fig.~\ref{fig:teste_01}.
The number of localized states in the energy region close to the Fermi level, highlighted by peaks in the DOS, is an increasing function of the twist angle. 
In particular, the in-plane ($\mathcal{T}_{12}$) and inter-plane ($\mathcal{T}_{13}$ and $\mathcal{T}_{14}$) transports, exhibit also a complex number of features in the corresponding energy range. 
A general mark is the highly anisotropic behavior of interplane transport for electrons coming from lead 1. In addition, a rich scenario of conductance suppression and energy filters is verified, for all transport channels.
 
In the case of $\theta =45^{\circ}$, the inter-plane transport favors $\mathcal{T}_{14}$ against $\mathcal{T}_{13}$. 
Moreover, a particular $50\%$ to $50\%$ transport emerges at the energy $E -E_{\rm F} \approx -0.28$~eV. 
Interestingly, the same phenomenon is verified for $\theta =15^{\circ}$ in smaller energy regions ($E -E_{\rm F} \approx -1.25$~eV), but in this case related to the $\mathcal{T}_{13}$-transmission (opposite direction). 
The use of TBZGNR systems as 50/50 electronic beam splitter, with the switch on and off controlled by varying the Fermi level by doping process \cite{Caio2016}, is extended here by the additional mechanism provided by swapping the twist angle $\theta$, which was shown to be experimentally manageable \cite{Wang2023}.
A remarkable aspect to be mentioned is the contrast between the soft evolution of the in-plane transmission and the inter-plane transmissions, outside the energy region of the central band, and the conductance fluctuations occurring in the later energy region. Also to be highlighted is the change from =$\mathcal{T}_{14}$ transmission to $\mathcal{T}_{13}$,  outside the central band range, as the twisted angle change from $60^{\circ}$ to $15^{\circ}$.

\begin{figure}[h!]
\centering
\includegraphics[width=0.98\linewidth]{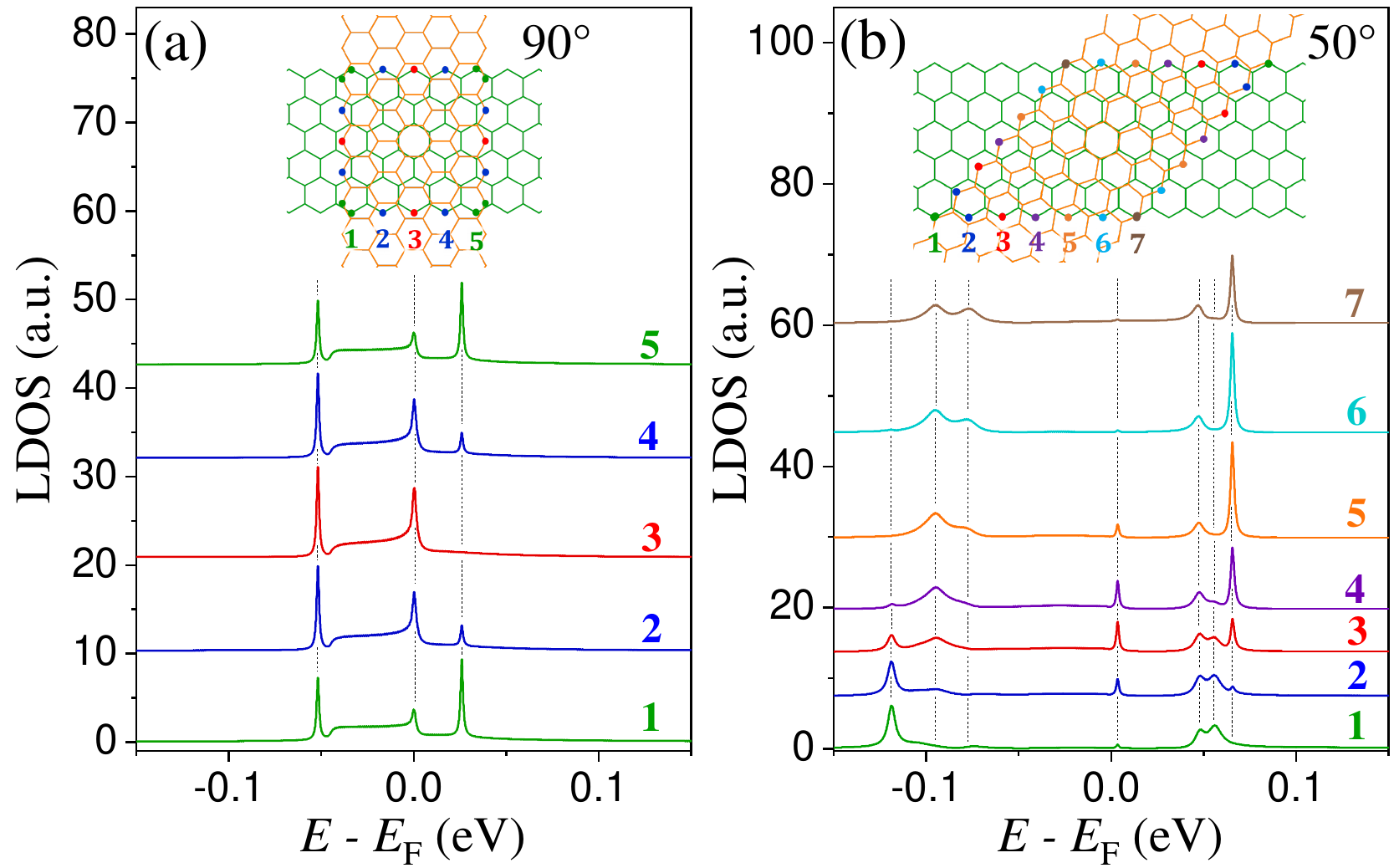}
\caption{
LDOS for the edge sites of $\theta = 90^{\circ}$ (model A) and $\theta = 50^{\circ}$.
Curves 1-5 in (a) and  1-7 in (b) correspond to the LDOS at atoms 1-5 and 1-7, indicate in the insets. 
Gray lines indicate the peak energy position.
}
\label{fig:Fig_angle0}
\end{figure}

To better understand the dependence of the electronic responses under changes in the stacking regions governed by the twist angle, we explore the local density of states of some particular atomic sites of the stacking regions. 
The local density of states at atomic positions layering at the limits of the staking region of TBZGNRs are shown in Figs.~\ref{fig:Fig_angle0}(a) and (b), considering $\theta = 90^{\circ}$ (model A) and $\theta = 50^{\circ}$, respectively. 
The colored curves correspond to sites 1-5 in Fig.~\ref{fig:Fig_angle0}(a) and 1-7 in Fig.~\ref{fig:Fig_angle0}(b), depicted in the insets.
The vertical gray lines in Fig. \ref{fig:Fig_angle0} indicate the positions of the LDOS peak energies. 
As the twisted angle decreases, the number of non-equivalent sites at the edges increases, as expected. 
Our results indicate that the number of LDOS peaks revealed in the LDOS calculation is the same as the number of non-equivalent sites at the stacking frontier. 
A further analysis involving TBZGNR with different twisted angles predicts the evolution of the LDOS peaks.

%--------------------------- F I G U R E ----------------------------------------
%-------------------------------------------------------------------------------

%--------------------------- F I G U R E ----------------------------------------
\begin{figure}[h!]
\centering
\includegraphics[width=0.95\linewidth]{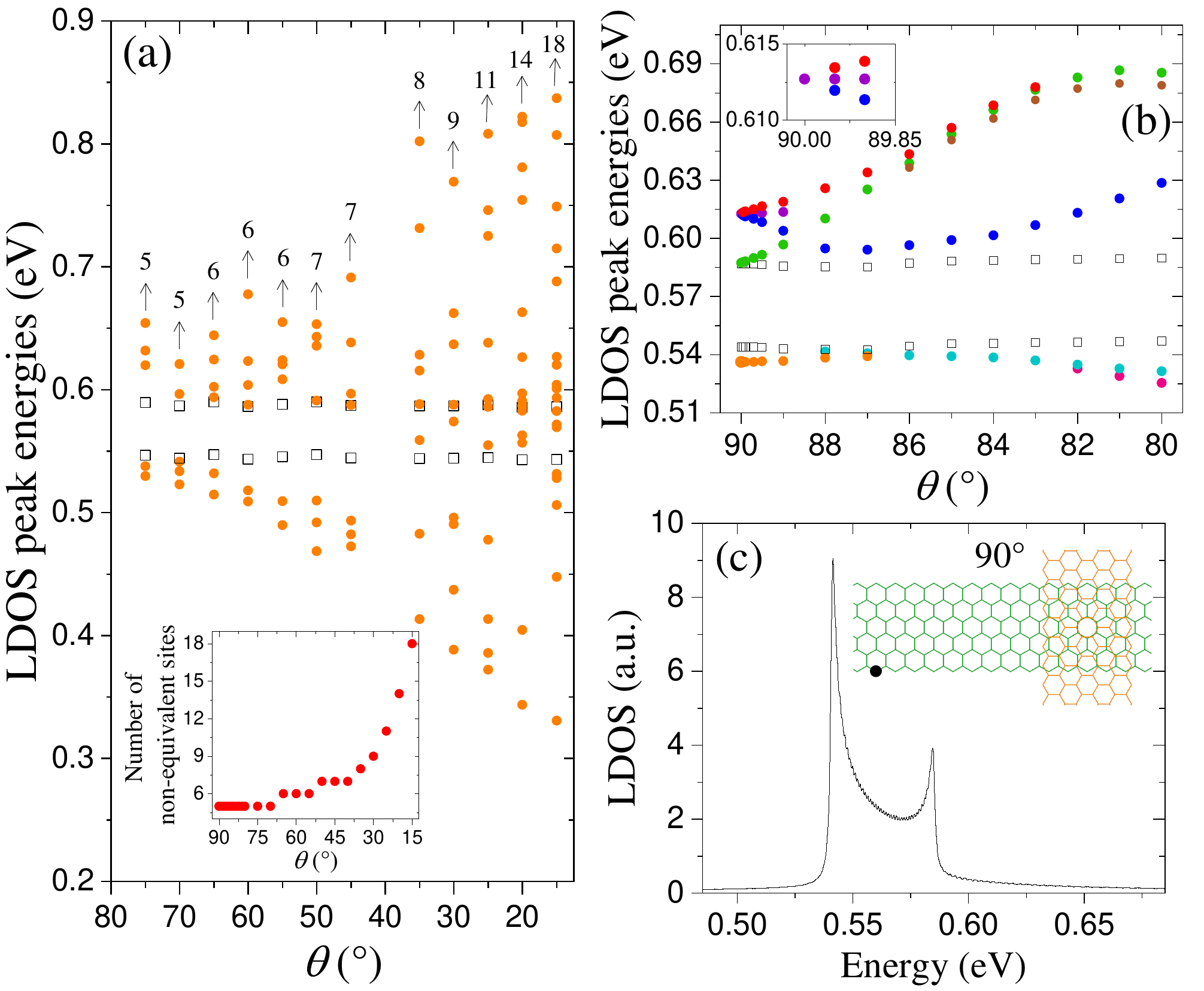}
\caption{
(a) LDOS peak energy spectra as a function of $\theta$ for $80^{\circ}\geq \theta \geq 15^{\circ}$.
Data labels at each angle correspond to the count of LDOS peaks.
The inset shows the number of non-equivalent sites as a function of $\theta$.
(b) Evolution of the LDOS peak energies as a function of the angle $\theta$ for $90^{\circ}\geq \theta \geq 80^{\circ}$.
Symbols with the same color represent the displacement of a particular LDOS peak.
Circle and square symbols correspond to  peaks of the LDOS for edge sites at the stacking zone and far from it, respectively. (C) LDOS of an edge state, identified in the inset by a black symbol, far from the stacking region.
}
\label{fig:Fig_angle}
\end{figure}
%-------------------------------------------------------------------------------

%--------------------------- F I G U R E ----------------------------------------
\begin{figure*}[t!]
%\begin{figure}[H]
\centering
\includegraphics[width=0.85\linewidth]{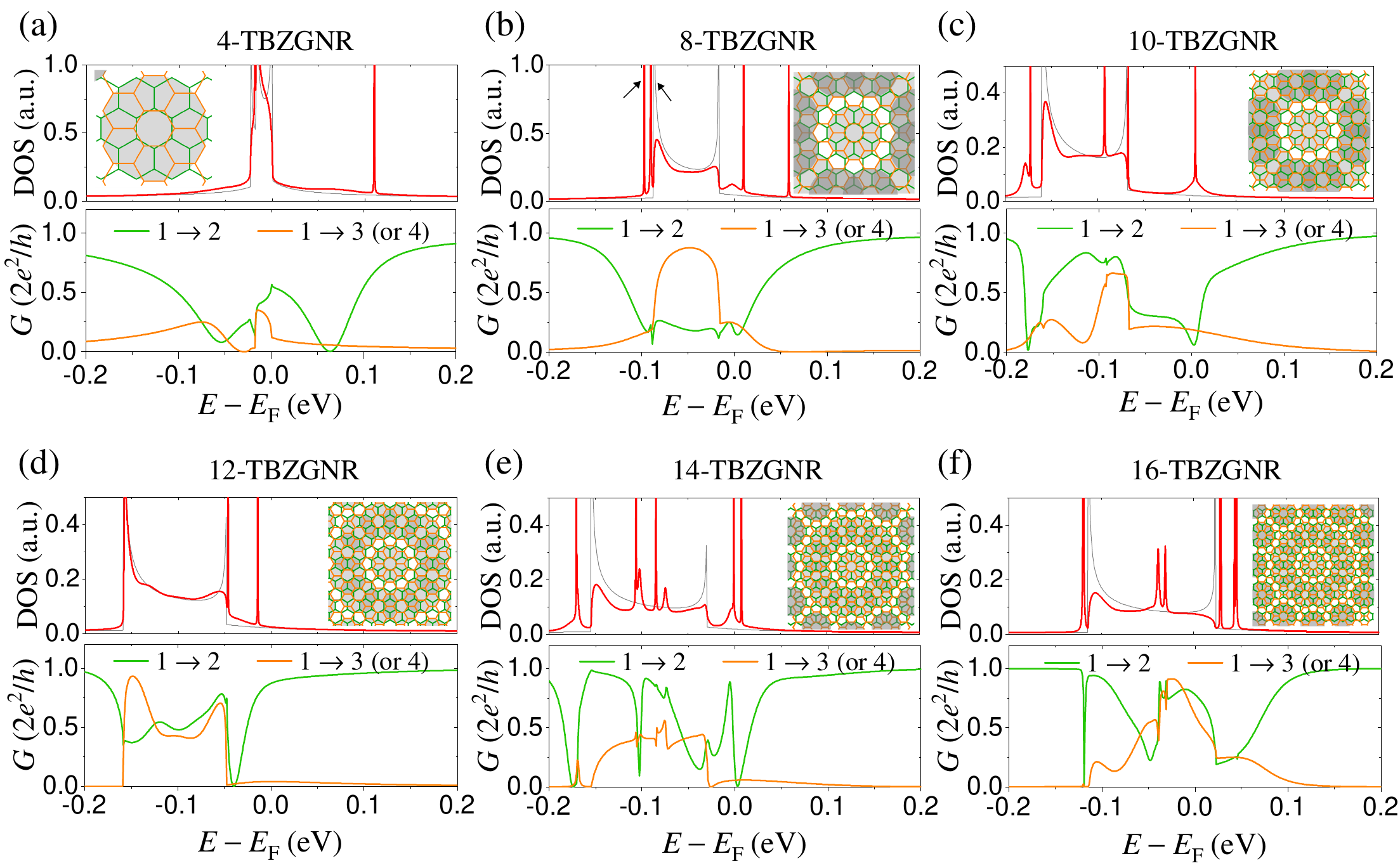}
\caption{DOS and conductance results for $90^{\circ}$ crossed zigzag-graphene nanoribbons: (a) 4-TBZGNR, (b) 8-TBZGNR, (c) 10-TBZGNR, and (d) 12-TBZGNR. Green (orange) curves denote $\mathcal{T}_{12}$ ($\mathcal{T}_{13}$ and $\mathcal{T}_{14}$). The 12-fold local-symmetry regions of each system are displayed in the insets.}
\label{fig:Fig_0x}
%\end{figure}
\end{figure*}

The dependence of the LDOS peak energies on the twist is presented in Fig. \ref{fig:Fig_angle}(a) for $80^{\circ}\geq \theta \geq 15^{\circ}$. 
The numerical data at each angle in the energy spectra indicate the number of LDOS peaks (orange disks) in the energy region of the central miniband. 
The number of non-equivalent sites for each TBZGNR configuration, as a function of $\theta$, is also shown in the inset. Our findings show a perfect correspondence between those features (LDOS peaks) and the non-equivalent site at the edges of the stacking region.  
The evolution of the DOS peak energies as a function of the small $\theta$ range,  $90^{\circ}\geq \theta \geq 80^{\circ}$, for which the number of non-equivalent sites is pinned and equal to 5, is exhibited in Fig. \ref{fig:Fig_angle}(b). 
Symbols of the same color represent the energy displacement of a particular LDOS peak as the angle decreases.  
The square symbols presented in both Figs~\ref{fig:Fig_angle}(a) and (b) correspond to LDOS peaks for an edge site (black dot in the inset) far from the stacking region, as shown in Fig. \ref{fig:Fig_angle}(c).

It is important to mention that the present results indicate that TBZGNRs offer a great number of opportunities to manipulate transport responses by changing the twisted angle and/or the Fermi energy of the system, which may be controlled by doped processes, for example.
The fact of being perfect 50/50 electronic-beam splitters, reveals them as natural candidates to be used as interferometer elements in nanodevices for varied angle configurations. Recent analysis has shown that crossed ZGNRs can further create a spin-polarizing scattering potential, working then as a spin-polarizing beam splitter \cite{Spin2022,Sanz2024}

%--------------------------- F I G U R E ----------------------------------------
\begin{figure}[h!]
\centering
\includegraphics[width=0.99\linewidth]{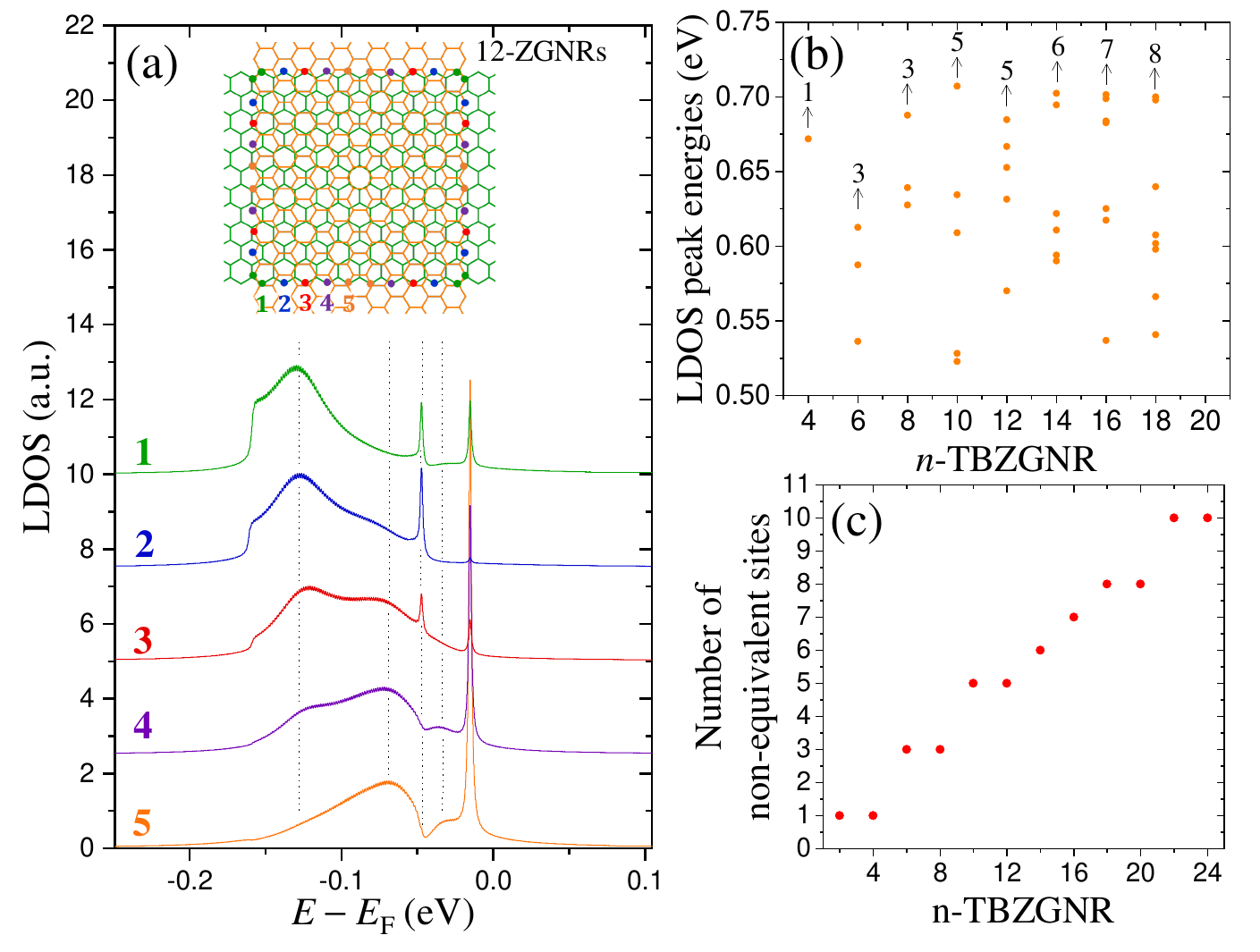}
\caption{(a) LDOS of the 5 different sites displayed at the contour boundary of the stacking region of a 12-TBZGNR, shown in the inset. Vertical lines denote the energy position of the LDOS peak. (b)
LDOS spectra for the peak energies as a function of the TBZGNR width. (c) Number of non-equivalent sites at the boundary for n-TBZGNR.}
\label{fig:Fig}
\end{figure}
%-------------------------------------------------------------------------------

\subsection{$90^{\circ}$- TBZGNR: varying the ribbon width}

We now explore the electronic and transport features of $90^{\circ}$ crossed ZGNRs described within model A, considering ribbons with increasing widths.
Figs.~\ref{fig:Fig_0x}(a)-(f) present the DOS and conductance of $90^{\circ}$-TBZGNRs of different widths. 
The 12-fold local symmetry regions for all the examples are depicted in the insets, highlighted with a light gray shadow. 
The geometric symmetries of such stacking configurations provide an equal transport response from lead 1 to terminals 3 and 4. 
In the case of 12-TBZGNR, a full dodecahedron symmetry is extended in the stacking region, as shown in the inset of Fig.~\ref{fig:Fig_0x}(d). It encloses an entire ring formed by superposed 12-fold local-symmetry regions. For wider nanoribbons, extra 12-fold rings emerge in the limiting stacking area. 
It is interesting to note that the interplane transports are essentially concentrated in the energy range of the original central peak of the single nanoribbon, defined by the two sharp peaks in the DOS, shown in gray curves in all pannels of Fig.~\ref{fig:Fig_0x}. 
Sharp transitions are observed at the emergence of the electronic tunneling of the bottom/top ribbons, when the interplane conductances ($\mathcal{T}_{13}$ and $\mathcal{T}_{14}$) achieve a finite value, close to one quantum conductance, for the 12-TBZGNR example, and lower values for $n=14$ and 16. Our findings show that although all the $90^{\circ}$-TBZGNR studied were designed within Model A, preserving both $x$- and $y$-mirror and $C_{2}$ symmetries, the resulting transport responses of each TBZGNR depend on the details of the stacking area. Actually, both in-plane transmission ($\mathcal{T}_{12}$) and interlayer transmissions ($\mathcal{T}_{13}$)  are quite sensitive to the arrangement of the dodecahedron figures in the pilled regions.

Similarly to what was discussed previously, the number of non-equivalent sites at the boundary of the stacking region was found to be equal to the LDOS peaks of such sites, as shown in Fig.~\ref{fig:Fig}(a), for the case of the 12-TBZGNR. This analysis was also performed for TBZGNRs with increasing widths, and the results are presented in Figs.~\ref{fig:Fig}(b) and (c).
While the LDOS peak spectra are shown in Fig.~\ref{fig:Fig}(b) as a function of the ribbon width, the number of non-equivalent sites at the boundary of the stacking region is depicted in Fig.~\ref{fig:Fig}(c), addressing this direct correspondence. It is important to mention that in counting the edge states of the stacking limited regions,
a distinction must be made in relation to particular energy values at which the feature (peak) is related to localized states outside the ribbon stacking, as 
discussed in Appendix B for the case of the 14-TBZGNR.
Our findings provided a detailed analysis on tunneling states emerging in the stacking regions, predicted previously in the experimental realization of the twisted bilayer zigzag-graphene nanoribbon junctions.

%%%%%%%%%%%%%%%%%%%%%%%%%%%%%%%%%%%%%%
\section{Conclusions}
\label{sec:conclusion}
%%%%%%%%%%%%%%%%%%%%%%%%%%%%%%%%%%%%%%
In the present study, we systematically investigated the electronic states and transport responses of twisted bilayer zigzag-graphene nanoribbons within a tight-binding approach. Our real-space model allows a realistic atomic description adopting energy hoppings through a considerable spatial neighborhood. 
Following experimental realization on such carbon-based nanostructures we discuss a variety of nanoribbon stacking configurations, including distinct symmetry aspects. Although our model does not include DFT or molecular dynamics lattice relaxation, it robustly captures the electronic behavior near the Fermi level. It was demonstrated \cite{Wang2023} that out-of-plane bending and lattice distortions due to van der Walls forces have negligible impact on the energy gap, ruling out bending and strain effects in the system.
Our findings show a direct connection between the number of non-equivalent sites on the edges of the stacking region and the localized states revealed by the analysis of local density of states close to the Fermi energy.

In-plane and interplane transports are studied across the 4-terminal scheme, resulting from the TBZGNR setup.  Conductance results reveal the possibility of getting full suppression of the transmission, current filters, favoring one lead against the others, and also interchanging of current dominant direction by changing the twisted angle. Additionally, we have observed the formation of 50/50 current filters for the in-plane and interplane channels at energy values quite dependent on the twisted angle.  By a controlled doping process these particular $50\%$ filters may be achieved.
The experimental ability to design diverse stacking options, allows the emergence of controllable tunneling electronic states fundamental for optical and electronic nanodevices.

%-----------------------------------------------------------------------
\acknowledgments

The authors thank the INCT de Nanomateriais de Carbono for providing support on the computational infrastructure. ABF and KJUV thanks the CNPq scholarships. AL thanks the FAPERJ under grant E-26/200.569/2023.
%-------------------------------------------------------------------------------

\appendix 
\setcounter{figure}{0}
\renewcommand{\thefigure}{A\arabic{figure}}

\section{Model C - Electronic Distribution}
\label{Model C}

\begin{figure}[h!]
\centering
\includegraphics[width=0.93\linewidth]{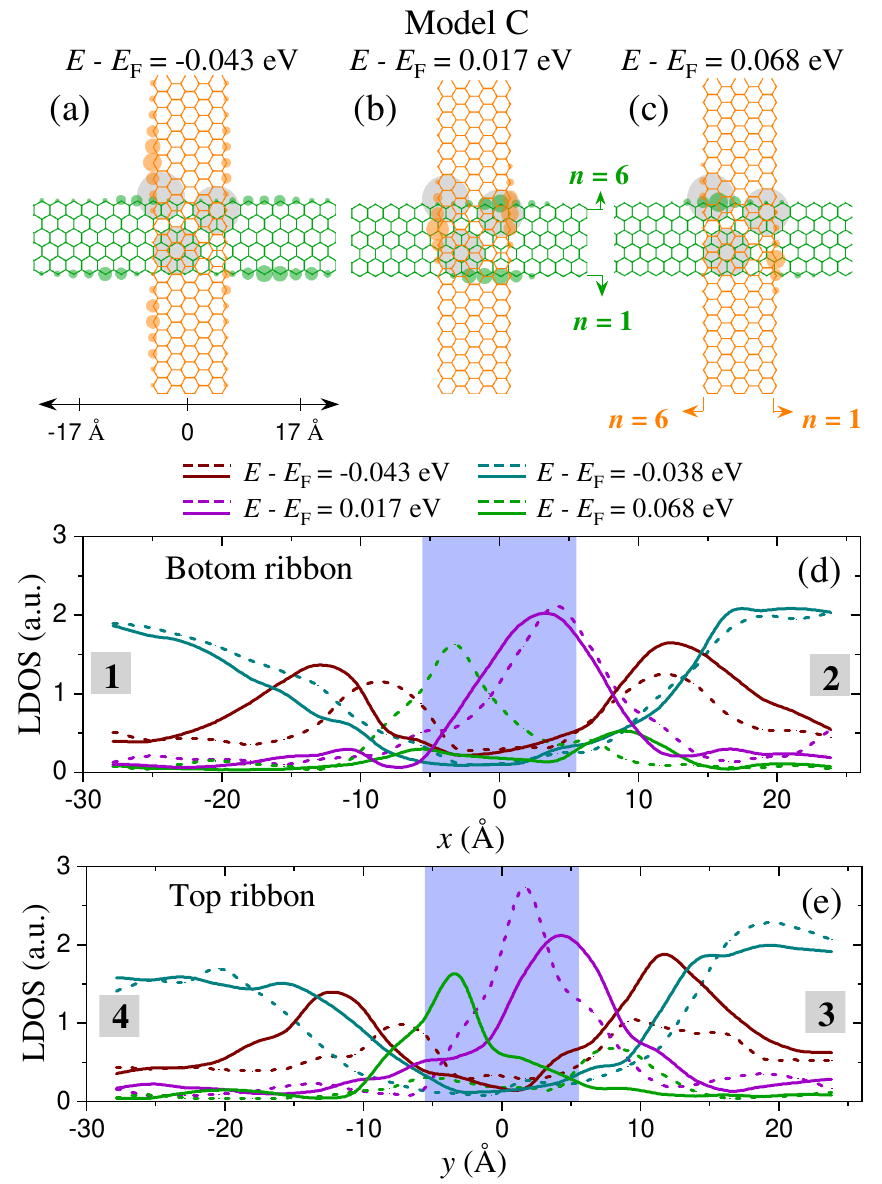}
\caption{Model C. LDOS at $E-E_{\rm F}=-0.043$~eV (a), $0.017$~eV (b), and $0.068$~eV (c) through the stacking region  and its next surroundings. 
LDOS at some energies at the (d) bottom and (e) top ribbon edge. Dashed and continuous lines denotes $n=1$ and $6$.}
\label{fig:Fig_A1}
\end{figure}

Here we explored the electronic distribution throughout the stacking region for Model C. Three energy cases are analyzed as shown in Figs.~\ref{fig:Fig_A1}(a), (b), and (c). 
They were also chosen as representative localized states marked by peaks in the corresponding DOS. Due to the lack of any symmetry, the LDOS are odd functions of the $x$ and $y$ positions, for the bottom and top ZGNRs. 

%--------------------------- F I G U R E ----------------------------------------
%-------------------------------------------------------------------------------

%\appendix 
\setcounter{figure}{0}
\renewcommand{\thefigure}{B\arabic{figure}}

\section{Local density of states}
\label{14-ZGNRs}

%--------------------------- F I G U R E ----------------------------------------
\begin{figure}[h!]
\centering
\includegraphics[width=0.78\linewidth]{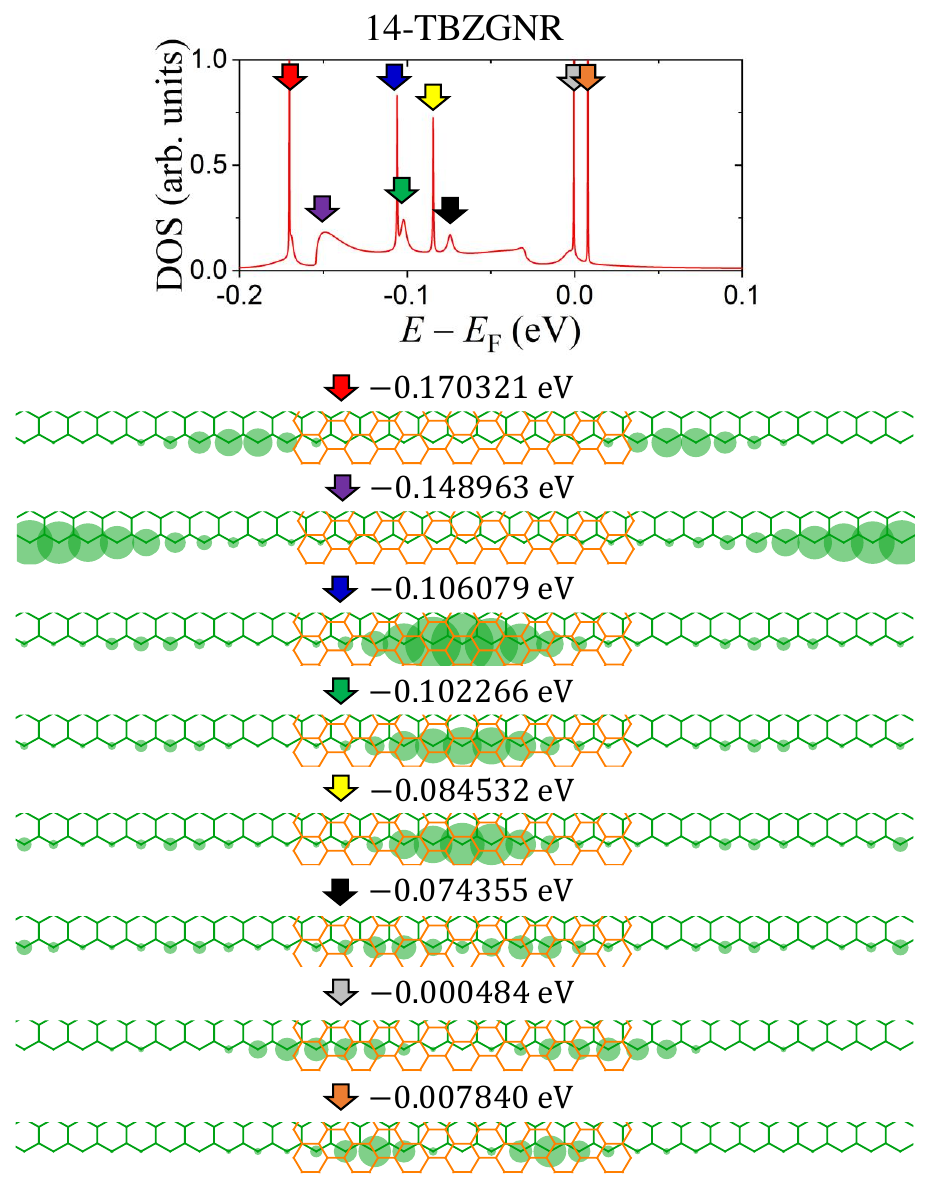}
\caption{
DOS as a function of energy for two crossed 14-TBZGNR. 
Colored arrows denote the energy positions of the DOS peaks. 
Bottom: LDOS projected on the bottom edge of the bottom ribbon through the stacking region and its next surroundings for the marked energies.
}
\label{fig:Fig_B1}
\end{figure}

Fig.~\ref{fig:Fig_B1} presents the LDOS of a 14-TBZGNR in an energy range closer to the Fermi energy.  
The particular energies corresponding to the LDOS features are marked with colored arrows. 
In the bottom part of Figure~\ref{fig:Fig_B1} we show the LDOS of each of these energies projected on the bottom edges of the bottom ribbon (green sites) through the stacking region and its next surroundings. 
Clearly, the result for $E-E_{\rm F}\approx -0.17$~eV does not correspond to a state located at the edge line of the stacking region. 
In addition, the state marked with the violet arrow is the ``linear bulk" state, a kind of revival of the isolated zigzag nanoribbon.  
All others are localized  states, centered in the boarding lines of the stacking region, as mentioned in the main text. 

\bibliography{cZGNRs,methods}
%-----------------------------------------------------------------------
\end{document}